\documentclass[pre,aps,10pt,twocolumn]{revtex4-1}
\usepackage{geometry}
\usepackage{graphicx}
\usepackage{amsfonts}
\usepackage{amsmath}
\usepackage{bm}
\usepackage{url}
\usepackage{color}
\usepackage{xfrac}
\usepackage{array,url,kantlipsum}

\newcommand{\vv}{{\bm v}}

\newcommand{\kv}{{\bm k}}

\newcommand{\Mm}{{\bm M}}

\definecolor{purple}{rgb}{0.5, 0.0, 0.8}
\definecolor{orange}{rgb}{0.9, 0.6, 0.0}

\newcommand{\Yes}{$\bullet$}
\newcommand{\No}{$-$}

\begin{document}
\title{Hyperuniformity and anti-hyperuniformity in one-dimensional substitution tilings} 

\author{Erdal C.~O\u{g}uz}
\affiliation{School of Mechanical Engineering and The Sackler Center for Computational Molecular and Materials Science, Tel Aviv University, Tel Aviv, Israel}
\author{Joshua E.~S.~Socolar}
\affiliation{Physics Department, Duke University, Durham, NC 27708, USA}
\author{Paul J.~Steinhardt}
\affiliation{Department of Physics \& Princeton Center for Theoretical Science, Princeton University,
Princeton, NJ 08544, USA}
\affiliation{Center for Cosmology and Particle Physics, Department of Physics, New York University,
New York, NY, 10003, USA}
\author{Salvatore Torquato}
\affiliation{Department of Chemistry, Department of Physics, Princeton Institute for the Science and Technology of Materials, and Program in Applied and Computational Mathematics, Princeton University, Princeton, New Jersey 08540, USA}

\date{\today}

\begin{abstract}
We consider the scaling properties characterizing the hyperuniformity (or anti-hyperuniformity) of long wavelength fluctuations in a broad class of one-dimensional substitution tilings.  We present a simple argument that predicts the exponent $\alpha$ governing the scaling of Fourier intensities at small wavenumbers, tilings with $\alpha>0$ being hyperuniform, and confirm with numerical computations that the predictions are accurate for quasiperiodic tilings, tilings with singular continuous spectra, and limit-periodic tilings.  Tilings with quasiperiodic or singular continuous spectra can be constructed with $\alpha$ arbitrarily close to any given value between $-1$ and $3$.  Limit-periodic tilings can be constructed with $\alpha$ between $-1$ and $1$ or with Fourier intensities that approach zero faster than any power law.
\end{abstract}

\maketitle 


\vspace{0.1in}
\section{Introduction}
Recent work has shown that spatial structures with density fluctuations weaker at long wavelengths than those of a typical random point set may have desirable physical properties, and such structures are said to be {\em hyperuniform}~\cite{To03a}. Crystals and quasicrystals are hyperuniform, as are a variety of disordered systems, including certain equilibrium structures, products of nonequilibrium self-assembly protocols, and fabricated metamaterials.  (For examples, see Refs.~\cite{Man2013,Haberko2013,Dre15,Torquato2015,Hex15,Castro-Lopez2017,Torquato2018}.)  One approach to generating point sets with nontrivial spatial fluctuations is to use substitution tilings as templates.  Our aim in this paper is to characterize the degree of hyperuniformity in such systems and thereby provide design principles for creating hyperuniform (or anti-hyperuniform) point sets with desired scaling properties.

Substitution tilings are self-similar, space-filling tilings generated by repeated application of a rule that replaces each of a finite set of tile types with scaled copies of some or all of the tiles in the set~\cite{Frank2008}.  We are interested in the properties of point sets formed by decorating each tile of the same type in the same way.  We consider here only one-dimensional (1D) tilings.  Although generalization to higher dimensions would be of great interest, the 1D case already reveals important conceptual features.

Substitution rules are known to produce a variety of structures with qualitatively different types of structure factors $S(k)$.  Some rules generate periodic or quasiperiodic tilings, in which case $S(k)$ consists of Bragg peaks on a reciprocal space lattice supported at sums and differences of a (small) set of basis wavevectors, which in the quasiperiodic case form a dense set.  Others produce limit-periodic structures consisting Bragg peaks located on a different type of dense set consisting of wavenumbers of the form $\pm k_0 n/p^m$, where $n$,  $m$, and $p$ are positive integers~\cite{Godreche1989,Baake2011}.  Still others produce structures for which $S(k)$ is singular at a dense set of points but does not consist of Bragg peaks~\cite{Bom86,Godreche1992,Baake2017}, for which $S(k)$ is absolutely continuous~\cite{Baake2012}, or for which the nature of the spectrum has not been clearly described.  The precise natures of the spectra are in many cases not fully understood.  

In this paper, we present a simple ansatz that predicts the scaling properties relevant for assessing the hyperuniformity (or anti-hyperuniformity) of 1D substitution tilings.  We illustrate the validity of the ansatz via numerical computations for a variety of example tilings that fall in different classes with respect to hyperuniformity measures.  We also delineate the full range of behaviors that can be obtained using the substitution construction method, including a novel class in which $Z(k)$ decays faster than any power.

Section~\ref{sec:alpha} reviews the definition of the scaling exponent $\alpha$ associated with both the integrated Fourier intensity $Z(k)$ and the variance $\sigma^2(R)$ in the number of points covered by a randomly placed interval of length $2R$.  We then review the classification of tilings based on the value of $\alpha$.  Section~\ref{sec:substitution} reviews the substitution method for creating tilings, using the well-known Fibonacci tiling as an illustrative example.   The substitution matrix $\Mm$ is defined and straightforward results for tile densities are derived.  Section~\ref{sec:scaling} presents a heuristic discussion of the link between density fluctuations in the tilings and the behaviors of $S(k)$ and $Z(k)$, which leads to a prediction for $\alpha$.  The prediction is shown to be accurate for example tilings of three qualitatively distinct types~\cite{Torquato2018}: strongly hyperuniform (Class~I), weakly hyperuniform (Class~III), and anti-hyperuniform.  
Section~\ref{sec:alpharange} shows, based on the heuristic theory, that the range of possible values of $\alpha$ produce by 1D substitution rules is $[-1,3]$ and that this interval is densely filled.  Section~\ref{sec:lp} considers substitutions that produce limit-periodic tilings.  Examples are presented of four distinct classes: logarithmic hyperuniform (Class~II), weakly hyperuniform (Class~III), anti-hyperuniform, and an anomalous class in which $Z(k)$ approaches zero faster than any power law.  Finally, Section~\ref{sec:discussion} provides a summary of the key results, including a table showing which types of tilings can exhibit the various classes of (anti-)hyperuniformity.

\section{Classes of hyperuniformity}
\label{sec:alpha}
For systems having a structure factor $S(\kv)$  that is a smooth function of the wavenumber $k$, $S(\kv)$ tends to zero as $k$ tends to zero \cite{To03a}, typically scaling as a power law
\begin{equation}
S(\kv) \sim k^{\alpha}.
\label{eqn:hyper}
\end{equation}
In one dimension, a unified treatment of standard cases with smooth $S(k)$ and quasicrystals with dense but discontinuous $S(k)$ is obtained by defining $\alpha$ in terms of the scaling of the integrated Fourier intensity
\begin{equation}
Z(k) = 2 \int_0^k  S(q) \, dq\,.
\label{Z}
\end{equation}
In both cases, $\alpha$ may be defined by the relation~\cite{Oguz2016}
\begin{equation}
Z(k) \sim k^{1+\alpha} \quad {\rm as} \; k \rightarrow 0\,.
\label{eqn:Z2}
\end{equation}
Systems with $\alpha>0$ have long wavelength spatial fluctuations that are suppressed compared to Poisson point sets and are said to be hyperuniform~\cite{To03a}.  Prototypical strongly hyperuniform systems (with $\alpha>1$) include crystals and quasicrystals.  We refer to systems with $\alpha<0$ as anti-hyperuniform~\cite{Torquato2018}.  Prototypical examples of anti-hyperuniformity include systems at thermal critical points.

An alternate measure of hyperuniformity is based on the local number variance of particles within a spherical observation window of radius $R$ (an interval of length $2R$ in the 1D case), denoted by $\sigma^2(R)$.  If $\sigma^2(R)$ grows more slowly than the window volume (proportional to $R$ in 1D) in the large-$R$ limit, the system is hyperuniform.  The scaling behavior of $\sigma^2(R)$ is closely related to the behavior of $Z(k)$ for small $k$~\cite{To03a,Oguz2016}.  For a general point configuration in one dimension with a well-defined average number density $\rho$,  $\sigma^2(R)$ can be expressed in terms of $S(k)$ and the Fourier transform ${\tilde \mu}(k;R)$ of a uniform density interval of length $2R$:
\begin{equation}
\sigma^2(R)=
2R\rho \Big[\frac{1}{2\pi} \int_{-\infty}^{\infty} S({k}) {\tilde \mu}(k;R) d{k}\Big] 
\label{eqn:local}
\end{equation}
with 
\begin{equation}
{\tilde \mu}(k;R)= 2\frac{\sin^2(k R)}{k^2 R}\,,
\label{eqn:alpha-k}
\end{equation}
where $\rho$ is the density.  (See Ref.~\cite{To03a} for the generalization to higher dimensions.)
One can express the number variance alternatively in terms of the integrated intensity~\cite{Oguz2016}:
\begin{equation}
\sigma^2(R)=
- 2R\rho\Bigg[\frac{1}{2\pi} \int_0^\infty  Z(k) 
\frac{\partial {\tilde{\mu}(k;R)}}{\partial k} dk \Bigg] \,.
\label{eqn:local-1}
\end{equation}

For any 1D system with a smooth or quasicrystalline structure factor, the scaling of $\sigma^2(R)$ for large $R$ is determined by $\alpha$ as follows~\cite{To03a,Za09,Torquato2018}:
\begin{equation} \label{eqn:alphanu}
\sigma^2(R) \sim \left\{\begin{array}{ll}
R^0, & \alpha > 1 \quad {\rm (Class\ I)} \\
\ln R, & \alpha = 1 \quad {\rm (Class\ II)}  \\
R^{1-\alpha}, &  \alpha < 1 \quad {\rm (Class\ III)} 
\end{array}\right.\,. 
\end{equation}
For hyperuniform systems, we have $\alpha>0$, and the distinct behaviors of $\sigma^2(R)$ define the three classes, which we refer to as strongly hyperuniform (Class~I), logarithmic hyperuniform (Class~II), and weakly hyperuniform (Class III).  Systems with $\alpha < 0$ are called ``anti-hyperuniform.''

The bounded number fluctuations of Class~I occur trivially for one-dimensional periodic point sets (crystals) and are also known to occur for certain quasicrystals, including the canonical Fibonacci tiling described below~\cite{Oguz2016}.  Other quasiperiodic point sets (not obtainable by substitution) are known to belong to Class~II~\cite{Kesten1966,Aubry1987,Oguz2016}.  

\section{Substitution tilings and the substitution matrix}
\label{sec:substitution}
A classic example of a substitution tiling is the one-dimensional Fibonacci tiling composed of two intervals (tiles) of length $L$ and $S$.  The tiling is generated by the rule 
\begin{equation} \label{eqn:fibsub}
L\rightarrow LS; \quad S\rightarrow L\,,
\end{equation}
which leads to a quasiperiodic sequence of $L$ and $S$ intervals.  
An important construct for characterizing the properties of the tiling is the {\em substitution matrix}
\begin{equation}
  \Mm = \left( \begin{array}{cc}  0 & 1 \\  1 & 1 \end{array}\right),
\end{equation}
which acts on the column vector $(N_S,N_L)$ to give the numbers of $S$ and $L$ tiles resulting from the substitution operation.

\begin{figure}
\includegraphics[width=\columnwidth]{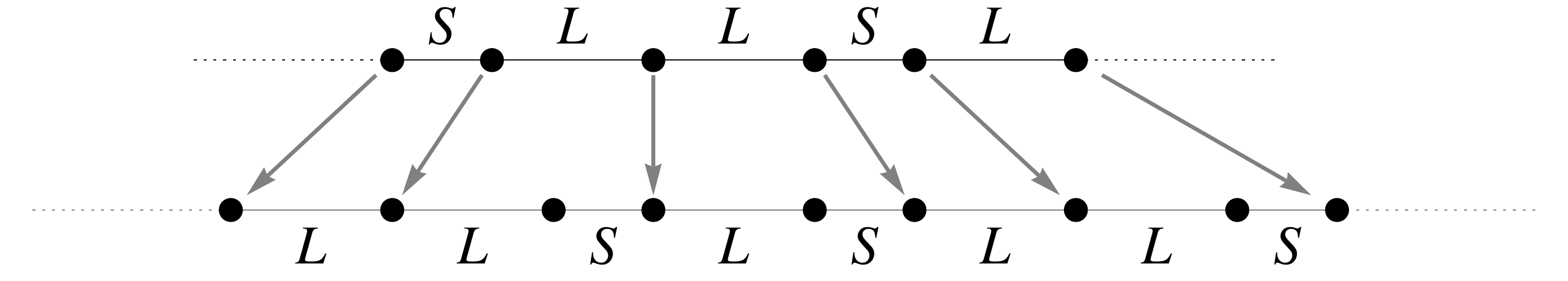}
\caption{The Fibonacci substitution rule.  The tiling on the upper line is uniformly stretched, then additional points are added to form tiles congruent to the originals.}
\label{fig:substitution}
\end{figure}

If the lengths $L$ and $S$ are chosen such that the ratio $L/S$ remains fixed, which in the present case requires $L/S = (1+\sqrt{5})/2\equiv\tau$, the substitution operation can be viewed as an affine stretching of the original tiling by a factor of $\tau$ followed by the division of each stretched $L$ tile into an $LS$ pair, as illustrated in Fig.~\ref{fig:substitution}.  
Given a finite sequence with $N_S$ tiles of length $S$ and $N_L$ tiles of length $L$, the numbers of $L$'s and $S$'s in the system after one iteration of the substitution rule is given by the action of the substitution matrix
on the column vector $(N_S,N_L)$.

More generally, substitution rules can be defined for systems with more than two tile types, leading to substitution matrices with dimension $D$ greater than 2.  We present explicit reasoning here only for the $D=2$ case.
A substitution rule for two tile types is characterized by a substitution matrix
\begin{equation} 
  \Mm = \left( \begin{array}{cc}  a & b \\  c & d \end{array}\right)\,.
\end{equation}
The associated rule may be the following:
\begin{equation}
S\rightarrow \underbrace{SS\ldots S}_{a}\underbrace{LL\ldots L}_{c}, \quad L\rightarrow \underbrace{SS\ldots S}_{b}\underbrace{LL\ldots L}_{d}\,,
\end{equation}
but different orderings of the tiles in the substituted strings are possible, and the choice can have dramatic effects.  Note, for example, that the rule
\begin{equation}
S\rightarrow SL, \quad L\rightarrow SLSL
\end{equation}
produces the periodic tiling $\ldots SLSLSL \ldots$, while the rule
\begin{equation}
S\rightarrow SL, \quad L\rightarrow SLLS
\end{equation}
produces the more complicated sequence discussed below in Section~\ref{sec:lp}.

Defining the substitution tiling requires assigning finite lengths to $S$ and $L$.  We let $\xi$ denote the length ratio $L/S$, and we consider only cases where the substitution rule preserves this ratio (i.e., $(bS + dL)/(aS + cL) = L/S$) so that the rule can be realized by affine stretching followed by subdivision.  This requires
\begin{equation}
\xi = \frac{d-a+\sqrt{(a-d)^2 + 4bc}}{2c}\,.
\end{equation}
For all discussions and plots below, we measure lengths in units of of the short tile length, $S$. 

The $SL$ sequence generated by a substitution rule is obtained by repeated application of that operation to some seed, which we will take to be a string containing $n_S$ short intervals and $n_L$ long ones.  We are interested in point sets formed by decorating each $L$ tile with $\ell$ points and each $S$ tile with $s$ points.  The total number of points at the $m^{th}$ iteration is  
\begin{equation}
{\cal N}_m = (s,\ell)\cdot \Mm^m \cdot (n_S,n_L),
\end{equation}
and the length of the tiling at the same step is  
\begin{equation}
{\cal X}_m = (1,\xi) \cdot \Mm^m \cdot (n_S,n_L).
\end{equation}
Let $\lambda_1$ and $\lambda_2$ be the eigenvalues of $\Mm$, with $\lambda_1$ being the largest, and let ${\vv}_1$ and ${\vv}_2$ be the associated eigenvectors.  We have
\begin{align}
\lambda_1  = a + c\,\xi\,; & \quad  
\lambda_2  = d - c\,\xi\,; \\
{\vv}_1  = \left(b/c,\, \xi\right)\,; & \quad
{\vv}_2  = \left(-\xi,\, 1\right)\,.
\end{align}
The unit vectors $(1,0)$ and $(0,1)$ may be expressed as follows:  
\begin{align}
(1,0) & = u(c\,{\vv}_1 - c\,\xi {\vv}_2)\,, \\
(0,1) & = u(c\,\xi {\vv}_1 +  b\, {\vv}_2)\,,
\end{align}
where $u = 1/(b+c\,\xi^2)$.  
We then have
\begin{eqnarray}
\Mm^m \cdot(n_S,n_L) &=& \Mm^m \cdot\big(n_S(1,0)+n_L(0,1)\big) \nonumber \\
 &=& u \bigg( \lambda_1^m \left(c\,n_S+ c\,\xi n_L\right){\vv}_1  \\
 &\ & \quad +  \lambda_2^m \left(-c\,\xi\,n_S+ b\,n_L\right){\vv}_2 \bigg)\,. \nonumber
\end{eqnarray}
The density of tile vertices after $m$ iterations, $\rho_m = {\cal N}_m/{\cal X}_m$, is thus 
\begin{equation}  \label{eqn:rhom}
 \rho_m  = \overline{\rho} + \left(\frac{\xi(s\xi-\ell)}{b+c\,\xi^2}\right)\left(\frac{c\,\xi  n_S - b\, n_L}{n_S+\xi n_L }\right) \left( \dfrac{\lambda_2}{\lambda_1} \right)^m,
\end{equation}
with $\overline{\rho} = (b s+c\ell\,\xi)/(b+c\,\xi^2)$, where we have used the fact that $(1,\xi)\cdot{\vv}_2 = 0$.

\section{Scaling properties of 1D substitution tilings}
\label{sec:scaling}
As long as the coefficient of $(\lambda_2/\lambda_1)^m$ in Eq.~(\ref{eqn:rhom}) does not vanish, the deviations of $\rho$ from $\overline{\rho}$ for portions of the tiling that are mapped into each other by substitution are related by
\begin{equation}
\delta\rho_{m+1} = \dfrac{\lambda_2}{\lambda_1}\delta\rho_m \,.
\label{eqn:deltarho}
\end{equation}
If the coefficient does vanish, which requires that $\xi$ be rational, the tiling may be periodic, but the ordering of the intervals in the seed becomes important.  We will revisit this point below.  For now we assume that the tiling is not periodic.

We make three conjectures regarding nonperiodic substitution tilings, supported, as we shall see, by numerical experiments.  The results are closely related to recently derived rigorous results~\cite{Baake2018b}.
\begin{description}
\item[Conjecture 1] 
We take Eq.~(\ref{eqn:deltarho}) to be the dominant behavior of density fluctuations throughout the system, not just for the special intervals that are directly related by substitution.  That is, we assume that there exists a characteristic amplitude of the density fluctuations at a given length scale after averaging over all intervals of that length, and that the $\delta\rho$ in Eq.~(\ref{eqn:deltarho}) can be interpreted as that characteristic amplitude.
\item[Conjecture 2]
We assume that the Fourier amplitudes $A(k)$ scale the same way as the density fluctuations at the corresponding length scale:
\begin{equation}
A(k/\lambda_1) = \frac{\lambda_2}{\lambda_1}A(k)\,.
\end{equation}
This implies the form
\begin{equation} \label{eqn:Ak}
A(k) \sim k^{(-\ln |\lambda_2/\lambda_1 |/ \ln |\lambda_1 |)} = k^{1-(\ln |\lambda_2 |/ \ln |\lambda_1 |)} \,.
\end{equation}
Squaring to get $S(k)$, we have
\begin{equation} \label{eqn:sofk}
S(k) \sim k^{(2 - 2 \ln |\lambda_2 |/ \ln |\lambda_1 |)}\,.
\end{equation}
This conjecture may not hold when interference effects are important, as in the case discussed in Sec.~\ref{sec:lp} below.
\item[Conjecture 3]
While $Z(k)$ is an integral of $S(k)$, the exponent must be calculated carefully when $S(k)$ consists of singular peaks.  
In the Fibonacci projection cases, the scaling of peak positions and intensities conspire to make $Z(k)$ scale with the same exponent as the envelope of $S(k)$~\cite{Oguz2016}.  We assume that this property carries over to substitution tilings with more than one eigenvalue greater than unity. Though the diffraction pattern is not made up of Bragg peaks~\cite{Bom86,Godreche1990}, we conjecture that it remains sufficiently singular for the relation to hold.  Thus we immediately obtain 
\begin{equation} \label{eqn:alpha}
\alpha = 1 - 2\left( \frac{\ln |\lambda_2 |}{\ln \lambda_1}\right)\,.
\end{equation}
\end{description}
Note that this calculation of the scaling exponent makes no reference to the distinction between substitutions with $|\lambda_2|<1$ and those with $|\lambda_2|>1$.  In the former case, $\lambda_1$ is a Pisot-Vijayaraghavan (PV) number, $S(k)$ consists of Bragg peaks, and $\sigma^2(R)$ remains bounded for all $R$.  In the latter case, the form of $S(k)$ is more complex~\cite{Bom86}, and quantities closely related to $\sigma^2(R)$, including the ``wandering exponent'' associated with lifts of the sequence onto a higher dimensional hypercubic lattice, are known to show nontrivial scaling exponents~\cite{Godreche1990}.

From Eq.~(\ref{eqn:alpha}), we see that the hyperuniformity condition $\alpha > 0$ requires $|\lambda_2| < \sqrt{\lambda_1}$.  Though the result was obtained for substitutions with only $D=2$ tile types, it holds for $D>2$ as well, so long as all ratios of tile lengths are preserved by the substitution rules; i.e., the dominant contribution to the long-wavelength fluctuations still scales like $|\lambda_2| / \lambda_1$. This distinction between hyperuniform and anti-hyperuniform substitution tilings thus divides the non-PV numbers into two classes that, to our knowledge, have not previously been identified as significantly different.  We note, for example, that the analysis presented in Ref.~\cite{Baake2018}, which treats substitution matrices of the form  $(0,n,1,1)$ and shows that they have singular continuous spectra (having no Bragg component or absolutely continuous component) for $n>2$, does not detect any qualitative difference between the cases $n=3$ and $n=5$.  The former case is hyperuniform, with $\lambda = (1/2)(1\pm\sqrt{13})$ and $\alpha \approx 0.37$, while the latter is anti-hyperuniform, with  $\lambda = (1/2)(1\pm\sqrt{21})$ and $\alpha \approx -0.14$.

\begin{figure}
  \centering
  \includegraphics[width=\columnwidth]{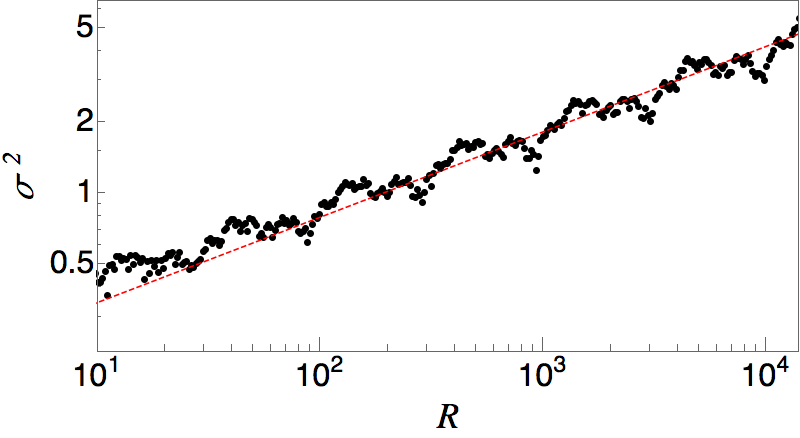}
  \caption{Log-log plot of the number variance (black dots) for a non-PV substitution tiling corresponding to $(a,b,c,d)=(0,2,2,2)$ decorated with points of equal weight at each tile boundary.  The variance was computed numerically for the tiling created by 11 iterations of the substitution on the initial seed $SL$.   The red dashed line has the predicted slope $1-\alpha \approx 0.36$.}
\label{fig:numvar_n02}
\end{figure}

For the Fibonacci case, we have  $\lambda_1 = \tau$ and $\lambda_2 = -1/\tau$, yielding $\alpha=3$, which agrees with the explicit calculation in Ref.~\cite{Oguz2016}).  
Considering $(a,b,c,d)$ of the form $(0,n,n,n)$ for arbitrary $n$, we find cases that allow explicit checks of our predictions for $\alpha$ for both hyperuniform and anti-hyperuniform systems.  We have $\lambda_1 = n\tau$ and $\lambda_2 = -n/\tau$, yielding
\begin{equation}
\alpha = 1-2\left( \frac{\ln n -  \ln \tau} {\ln n + \ln \tau}\right)\,.
\label{eq:exp_n}
\end{equation}
For $n\ge 2$, the system, the presence of more than one eigenvalue with magnitude greater than unity gives rise to more complex spectral features, possibly including a singular continuous component.
For $2 \le n \le 4$, our calculation predicts $0 < \alpha < 1$ and hence $\sigma^2(R) \sim R^{1-\alpha}$. 
We numerically verify the latter result for $n=2$ using a set of 954,369 points generated by 12 iterations of the substitution tiling, where the decoration consists of placing one point at the rightmost edge of each tile (with $s=\ell=1$).  Fig.~\ref{fig:numvar_n02} shows the computed number variance.  For each point, a window of length $2R$ is moved continuously along the sequence and averages are computed by weighting the number of points in the window by the interval length over which that number does not change.  A regression analysis yields $\sigma^2 (R) \sim R^{0.36}$, in close agreement with the predicted exponent from Eq.~(\ref{eqn:sofk}): $1-\alpha = 2 (\ln 2 -  \ln \tau) / (\ln 2 + \ln \tau) \approx 0.36094$.  

\begin{figure}
  \centering
  \includegraphics[width=\columnwidth]{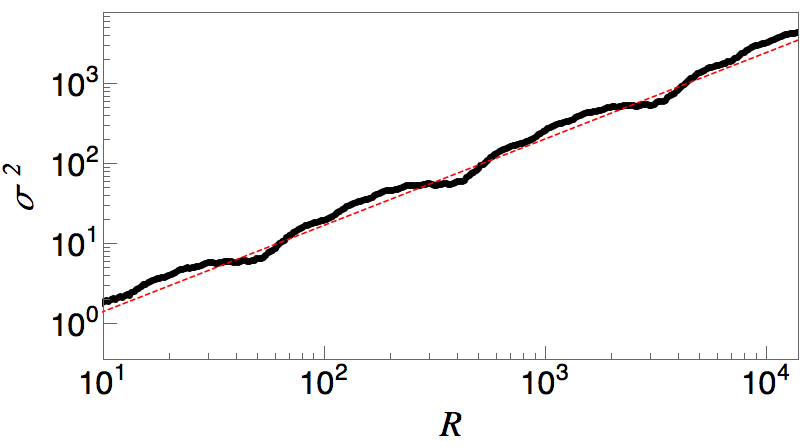}
  \caption{Log-log plot of the number variance (black dots) for an anti-hyperuniform substitution tiling corresponding to $(a,b,c,d)=(0,5,5,5)$ decorated with a points of equal weight at each tile boundary.   The variance was computed numerically for the tiling created by 6 iterations of the substitution on the initial seed $SL$.  The red dashed line has the predicted slope $1-\alpha \approx 1.08$.}
\label{fig:numvar_n05}
\end{figure}

For $n \ge 5$, the calculated value of $\alpha$ is negative, approaching $-1$ as $n$ approaches infinity.  The point set is therefore anti-hyperuniform; it contains density fluctuations at long wavelengths that are stronger than a those of a Poisson point set.  For $n=5$, we have $\alpha = -0.0793\ldots$.  Fig.~\ref{fig:numvar_n05} shows a log-log plot of the computed number variance along with the line corresponding to $\sigma^2 (R) \sim R^{1-\alpha}$.  Again the agreement between the numerical result and the predicted value is quite good.
Intuition derived from theories based on nonsingular forms of $S(k)$ suggest that a negative value of $\alpha$ should be associated with a divergence in $S(k)$ for small $k$, though it remains true that $Z(k)$ converges to zero for $\alpha > -1$.   For singular spectra, the envelope of $S(k)$ scales like $Z(k)$, and we do not expect any dramatic change in the behavior of $S(k)$  as $\alpha$ crosses from positive (hyperuniform) to negative (anti-hyperuniform).   The theories presented in  Refs.~\cite{Baake2017} and~\cite{Godreche1990} may provide a path to the computation of scaling properties of $S(k)$ in these cases.  It is worth noting, however, that the various classes of behavior can be realized by substitutions that produce limit-periodic tilings with $S(k)$ consisting entirely of Bragg peaks with no singular-continuous component, as shown in Section~\ref{sec:lp} below.

For rules that yield rational values of the length ratio $\xi$, the coefficient of $(\lambda_2/\lambda_1)^m$ in Eq.~(\ref{eqn:rhom}) can vanish for appropriate choices of $n_S$ and $n_L$, suggesting that there are no fluctuations about the average density that scale with wavelength.  This reflects the fact that the sequence of intervals associated with the substitutions can be chosen to generate a periodic pattern.  (A simple example is $S\rightarrow L$ and $L\rightarrow SLS$, which generates the periodic sequence $\ldots SLSLSL\ldots$, with $\xi = 2$, $\lambda_1=2$, and $\lambda_2=-1$.)  For such cases, $S(k)$ is identically 0 for all $k$ smaller than the reciprocal lattice basis vector.  For other interval sequence choices corresponding to the same $\Mm$, the tiling can be limit-periodic, and we would expect the scaling to be given by applying the above considerations with generic choices of the ordering, which would yield $\alpha=1$ and therefore a logarithmic scaling of $\sigma^2(R)$.  This case is presented in more detail in Section~\ref{sec:lp} below, and the logarithmic scaling is confirmed.  

\section{Achievable values of {\boldmath\large $\alpha$}}
\label{sec:alpharange}
Beyond establishing that substitution tilings exist for each hyperuniformity class, it is natural to ask whether any desired value of $\alpha$ can be realized by this construction method.  Here we show that if $\Mm$ is full rank, $\alpha$ always lies between $-1$ and $3$.   

First, note that the maximum value of $|\lambda_2/\lambda_1|$ is 1, by definition, which sets the lower bound on $\alpha$ via Eq.~(\ref{eqn:alpha}).  The upper bound on $\alpha$ is obtained when $|\lambda_2|$ is as small as possible, but there is a limit on how small this can be.  The product of the eigenvalues of $\Mm$ is equal to $\det \Mm$, so $|\lambda_2|$ cannot be smaller than $(|\det\Mm| / \lambda_1)^{\frac{1}{(D-1)}}$.  But $|\det\Mm|$ is an integer, and the smallest value nonzero value it can take is $1$.  (The case $\lambda_2 = 0$ is discussed in Section~\ref{sec:lp} below.  For $D\ge 3$, one can have $\det\Mm = 0$ with nonzero $\lambda_2$.  The analysis of such cases is beyond our present scope.) Hence we have 
\begin{equation}
|\lambda_2| \ge  \lambda_1^{-1/(D-1)} \,,
\end{equation}
implying
\begin{equation}
\alpha = 1 - 2 \frac{\ln|\lambda_2|}{\ln\lambda_1} \le \frac{D+1}{D-1}\,.
\end{equation}
Thus the maximum value of $\alpha$ obtainable by this construction method is $3$, which can occur for $D=2$, as in the Fibonacci case.

The family of substitutions considered in Section~\ref{sec:scaling} above produces a discrete set of values of $\alpha$ ranging from $-1$ to $3$.  
By considering two additional families, we can show that the possible values of $\alpha$ densely fill this interval.  For 
\begin{equation}
\Mm = \left(\begin{array}{cc} a & 0 \\ c & d\end{array}\right)
\end{equation}
with $d>a+1$ and $2c<(d-a)$, we have $\lambda_1 = a$ and $\lambda_2 = d$.  Note that $\xi = (d-a)/c$ is rational here;  we assume that the substitution sequences for the two tiles are chosen so as to avoid periodicity.  We have
\begin{equation}
\alpha = 1 - 2 \frac{\ln a}{\ln d}\,.
\end{equation}
For fixed $a$, $d$ can range from $a+2$ to $\infty$.  As $d$ approaches infinity, $\alpha$ approaches $1$.  For $d = a+2$, as $a$ approaches infinity, $\alpha$ approaches $-1$.  For sufficiently large $d$, the values of $a$ between $1$ and $d-2$ yield an arbitrarily dense set of $\alpha$'s between $-1$ and $1$.

Another class of $\Mm$'s produces $\alpha$'s between $1$ and $3$.  For
\begin{equation}
\Mm = \left(\begin{array}{cc} 0 & b \\ b & n\,b \end{array}\right)
\end{equation}
with $n>b$, we have 
\begin{align}
\xi & = \frac{1}{2}(n + \sqrt{n^2+4})\,,\\
\lambda_{1,2} & =  \frac{b}{2}(n \pm \sqrt{n^2+4})\,,\\
\end{align}
We thus obtain 
\begin{equation}
\alpha = 1 - 2 \frac{\ln b - \ln 2 + \ln (\sqrt{n^2+4}-n)}{\ln b - \ln 2 + \ln (\sqrt{n^2+4}+n)}\,.
\end{equation}
For large $n$, we have 
\begin{equation}
\alpha \approx 1 - 2 \frac{\ln b - 2\ln 2 - \ln n}{\ln b + \ln n}\,,
\end{equation} 
which approaches $3$ for $b \ll n$ and approaches $1$ for $b=n$.  By making $n$ as large as desired, the values of $b$ between $1$ and $n$ give $\alpha$'s that fill the interval between $1$ and $3$ with arbitrarily high density.

\section{Limit-periodic tilings}
\label{sec:lp}
For a limit-periodic tiling, the set of tiles is a union of periodic patterns with ever increasing lattice constants of the form $a p^n$, where $p$ is an integer and $n$ runs over all positive definite integers~\cite{Godreche1989,Baake2011,Socolar2011}.
We show here that there exist limit-periodic tilings of four hyperuniformity classes:  logarithmic (Class II), weakly hyperuniform (Class III), anti-hyperuniform, and an anomalous case in which $Z(k)$ decays to zero faster than any power law as $k$ goes to zero.
The latter corresponds to a rule for which $\det\Mm = 0$ (and $\lambda_2 = 0$), in which case $\alpha$ is not well defined. 
The existence of anti-hyperuniform limit-periodic tilings shows that anti-hyperuniformity does not require exotic singularities in $S(k)$ for small $k$.  Generally, it requires only that $Z(k)$ grows sub-linearly with $k$.

\subsection{The logarithmic case ($\alpha = 1$)} 
The rule $L\rightarrow LSS$, $S\rightarrow L$ with $S=1$ and $L=2$ yields the well-known ``period doubling'' limit-periodic tiling.  The eigenvalues of the substitution matrix are $\lambda_1 = 2$ and $\lambda_2 = -1$, leading to the prediction $\alpha=1$ and therefore quadratic scaling of $Z(k)$ and logarithmic scaling of $\sigma^2(r)$.  
Numerical results for $\sigma^2(R)$ are in good agreement with this prediction~\cite{Torquato2018b}. 
\begin{figure}
  \centering
  \includegraphics[width=\columnwidth]{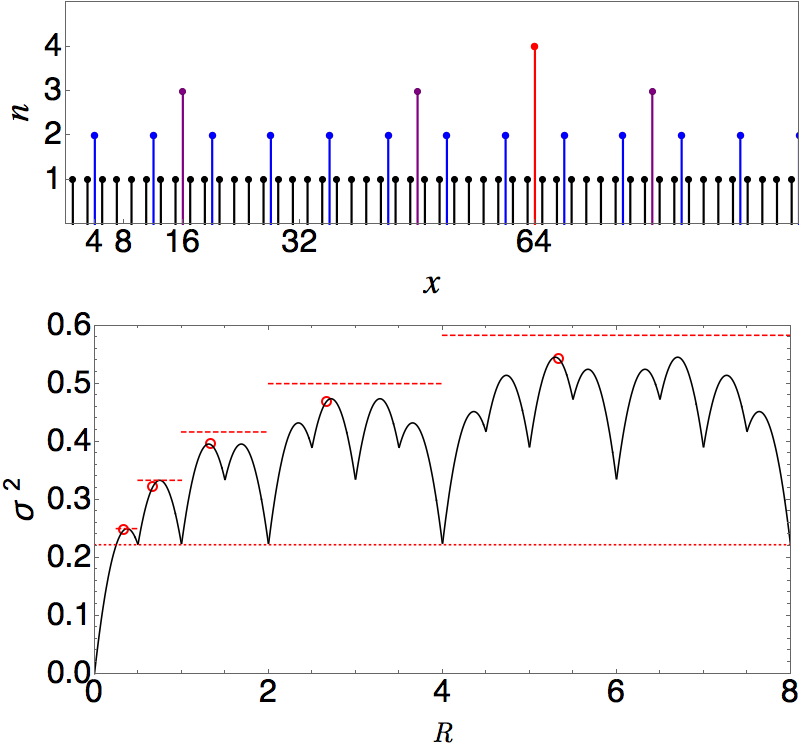}
  \caption{The $\alpha=1$ (period doubling) limit-periodic tiling.  Top: the tile boundaries with each point plotted at a height corresponding to the value of $n$ for the sublattice it belongs to.  Bottom: Plot of the number variance.  The horizontal dotted line marks $\sigma^2=2/9$, which is obtained for every $R$ of the form $2^n$ with integer $n\geq-1$.  The dashed lines indicate upper bounds, and the open circles are analytically calculated values for $R = 2^n/3$.  See text for details.}
\label{fig:numvar_lpa1}
\end{figure}
In fact, one can show explicitly via direct calculation of $\sigma^2(R)$ that the scaling is logarithmic.  The calculation outlined in the appendix shows that
\begin{equation} \label{eqn:s2sum}
\sigma^2(R) = \frac{1}{3}\sum_{n=0}^\infty \left\{\frac{w}{2^n}\right\}\left(1-\left\{\frac{w}{2^n}\right\}\right)\,,
\end{equation}
where $w = 2R$ and $\{x\}$ denotes the fractional part of $x$.  From this it follows that for $R = 2^{n-1}/3$ with $n\geq 1$ we have
\begin{equation}
\sigma^2(R) = \frac{2}{27} \left( \frac{13}{3} + n \right)\,,
\end{equation}
demonstrating clear logarithmic growth for this special sequence of $R$ values.  One can also derive an upper bound over the interval $2^{n-1}<R\leq 2^n$ by assuming that the summand in Eq.~(\ref{eqn:s2sum}) takes its maximum possible value on the intervals $(0,1/2], (1/2,1], (2^{m-1},2^{m}]$, for $m\leq n$, and maximizing the possible sum of the exponentially decaying remaining contributions.  The result is
\begin{equation}
\sigma^2(R) < \frac{1}{4} \left(1 + \frac{n}{3}\right) \quad {\rm for\ } 2^{n-1}<R\leq 2^n\,.
\end{equation}
This upper bound also grows logarithmically and is shown as a series of dashed lines in Fig.~\ref{fig:numvar_lpa1}.

\begin{figure}
  \centering
  \includegraphics[width=\columnwidth]{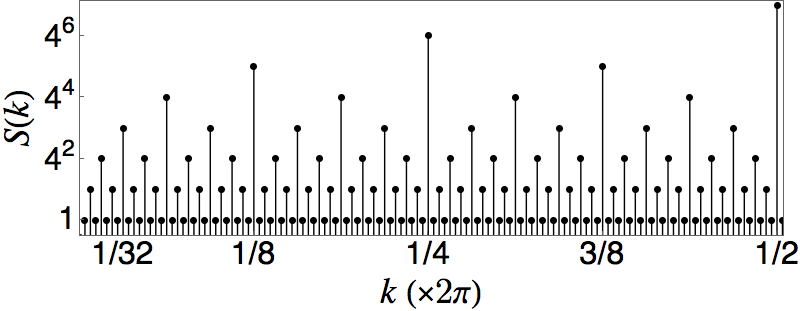}
  \includegraphics[width=\columnwidth]{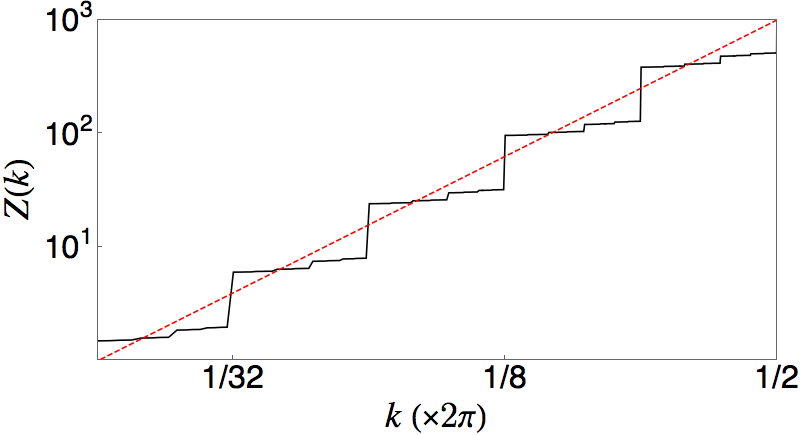}
  \caption{The $\alpha=1$ limit-periodic tiling.  Top: a logarithmic plot of the analytically computed $S(k)$ (arbitrarily scaled) including $k_{mn}$ with $n\leq 3$.  Bottom: a log-log plot of $Z(k)$ computed numerically from $S(k)$.   The dashed red line shows the expected quadratic scaling law.}
\label{fig:lpdoubling}
\end{figure}
It is instructive to carry out a more detailed analysis of  $S(k)$ for this particularly simple case as well.  (See also Ref.~\cite{Torquato2018b}.)  The tiling generated by applying the substitution rule repeatedly to a single $L$ with its left edge at $x=1$ consists of points located at positions $4^{\ell} (2j+1)$, where $\ell$ and $j$ range over all positive integers (including zero).  The structure factor therefore consists of peaks at $k_{mn} = 2\pi m/(a p^n)$, with $a=2$ and $p=4$, for arbitrarily large $n$ and all integer $m$.  For $m$ not a multiple of $4^{n-1}$, the peak at $k_{mn}$ gets nonzero contributions only from the lattices with $\ell \geq n$.
These can be summed as follows:
\begin{eqnarray}
S(k_{mn}) & = & \lim_ {N\rightarrow\infty} \bigg|  \sum_{\ell = n}^{\infty} \left( \frac{1}{2\times 4^{\ell}}\right) \nonumber \\
  \ & \ & \times \frac{1}{N} \sum_{j=0}^{N-1} \exp\left(\frac{2\pi i m 4^{\ell}(2j+1)}{2\times4^n}\right) \bigg|^2 \\
\ & = & \left(\frac{1}{9 \times 4^{2n}}\right) 4^{\rm{mod}_2(m+1)}\,,
\end{eqnarray}
where the factor of $1/(2\times 4^{\ell})$ in the first line is the density of the sublattice with that lattice constant.  Applying this reasoning to each value of $n$ gives a result that can be compactly expressed as
\begin{equation}
S(k_{m\nu}) = \left[\frac{{\rm GCD}\left(2^{\nu},m\right)}{3 \times 4^{\nu}} \right]^2\,,
\end{equation}
where $\nu$ is an arbitrarily large integer, ${\rm GCD}()$ is the greatest common denominator function, and $m$ can now take any positive integer value.
Figure~\ref{fig:lpdoubling} shows plots of $S(k)$ and $Z(k)$ for this tiling.  (See also Ref.~\cite{Torquato2018b} for an explicit expression for $Z(k)$ and proof of the quadratic scaling.)  Note that the apparent repeating unit in the plot of $Z(k)$ spans only a factor of 2, even though the scaling factor for the lattice constants is 4.  A similar effect occurs in the Poisson and anti-hyperuniform cases below.  In the present case, the construction in the appendix showing that the density can be expressed using lattice constants $1/2^n$ explains the origin of the effect.

\subsection{A Poisson scaling example ($\alpha = 0$) \\ and weak hyperuniformity ($0<\alpha<1$)}
\begin{figure}
  \centering
  \includegraphics[width=\columnwidth]{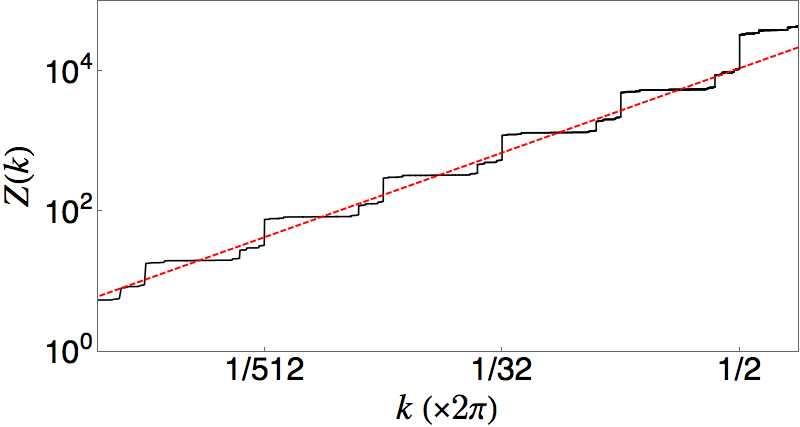}
  \includegraphics[width=\columnwidth]{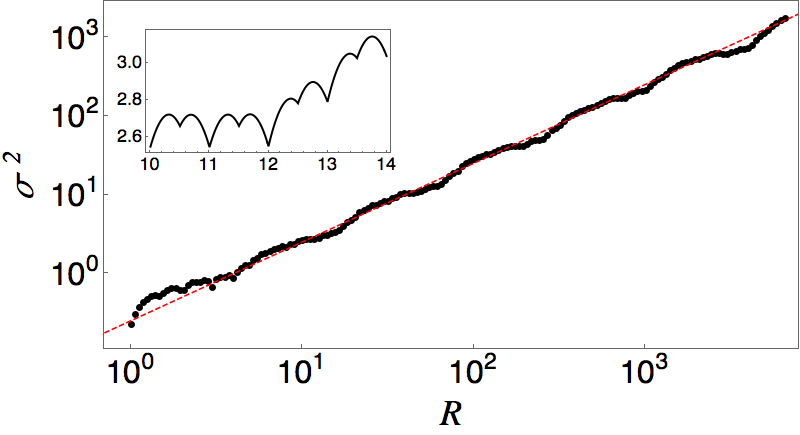}
  \caption{Comparison of direct computation of $Z(k)$ and $\sigma^2(R)$ with the predicted scaling laws for a limit-periodic tiling with $\alpha = 0$.  The dashed red lines show the expected linear scaling laws.  The inset shows the piecewise parabolic behavior of $\sigma^2(R)$ over a small span of $R$ values.}
\label{fig:lppoisson}
\end{figure}

The substitution rule 
\begin{equation}
S\rightarrow LL, \quad L\rightarrow LLSSSS
\label{eqn:lprule}
\end{equation}
with $S=1$ and $L=2$ produces a limit-periodic tiling with $a=2$ and $p = 16$.  
Equation~(\ref{eqn:alpha}) yields $\alpha = 0$, which is the value corresponding to a Poisson system.
Figure~\ref{fig:lppoisson} shows the result of direct computations of $Z(k)$ including all of the Bragg peaks at $k = 2\pi n/ (a p^3)$ and of $\sigma^2(R)$.  Values of $\sigma^2$ were computed from a sequence of $21,889$ points obtained by seven iterations of the substitution rule on an initial $L$ tile.  For each point, a window of length $2R$ is moved continuously along the sequence for the computation of the averages.

Limit-periodic examples of weak hyperuniformity (Class~III) are afforded by substitutions of the form 
\begin{equation}
\Mm = \left( \begin{array}{cc}  0 & 2n \\  2 & 2(n-1) \end{array}\right)\,,
\end{equation}
with $n\geq 3$ with  $L/S=n$, which yields 
\begin{eqnarray}
\alpha  & = & \frac{\ln n - \ln 2}{\ln n + \ln 2} \\
\ & = & \{0.226294, 1/3, 0.39794, 0.442114, \ldots\}\,.
\end{eqnarray}
 
\subsection{Anti-hyperuniformity ($\alpha < 0$)}
\begin{figure}
  \centering
  \includegraphics[width=\columnwidth]{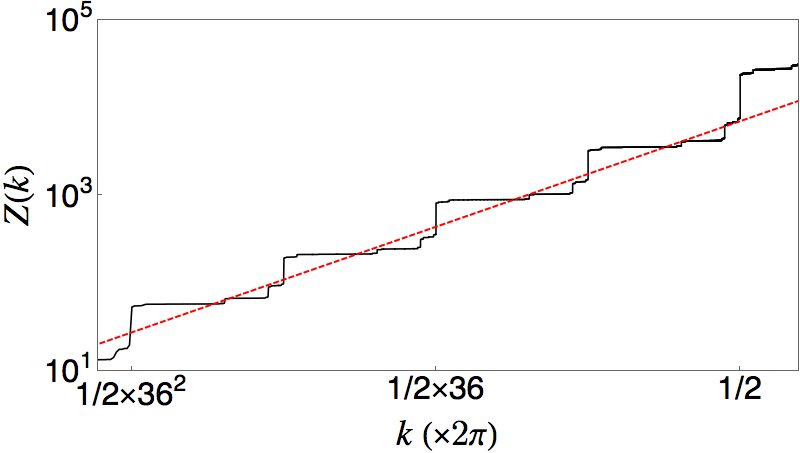}
  \caption{Comparison of direct computation of $Z(k)$ with the predicted scaling law for a limit-periodic tiling with $\alpha =  -0.226294\ldots$.  The dashed red line shows the expected scaling law with slope $1+\alpha$.}
\label{fig:lpantihyper}
\end{figure}

The substitution rule 
\begin{equation}
S\rightarrow LLL, \quad L\rightarrow LLLSSSSSS
\label{eqn:lprule}
\end{equation}
with $S=1$ and $L=2$ produces a limit-periodic tiling with $a=2$ and $p = 36$.  
Equation~(\ref{eqn:alpha}) yields 
\begin{equation}
\alpha = 1-2\frac{\ln 3}{\ln 6} = -0.226294\ldots\,,
\end{equation}
 which indicates anti-hyperuniform fluctuations.
Figure~\ref{fig:lpantihyper} shows the result of a direct computation of $Z(k)$ including all of the Bragg peaks at $k = 2\pi n/ (a p^3)$.

More generally, substitution matrices of the form 
\begin{equation}
\Mm = \left( \begin{array}{cc}  0 & 2 n \\   n & n \end{array}\right)
\end{equation}
with $n\geq 3$ and $L/S=2$ yield limit-periodic tilings anti-hyperuniform tilings with 
\begin{eqnarray}
\alpha & = & 1 - 2\frac{\ln n}{\ln 2n} = \frac{\ln 2 - \ln n}{\ln 2 + \ln n}\\
\ & = &\{-0.226294, -1/3, -0.39794, \ldots\}\,.
\end{eqnarray}

\subsection{A $\lambda_2 = 0$ case ($\alpha$ undefined)}
A special class of tilings is derived from substitution matrices of dimension $D=2$ that have $\lambda_2 = 0$ (and hence $\det\Mm = 0$).  Such rules can produce periodic tilings, limit-periodic ones, or more complex structures.  The criteria for limit-periodicity can be obtained by analyzing constant-length substitution rules in which each $L$ is considered to be made up of two tiles of unit length: $L = AB$.  If the induced substitution rule on $S$, $A$, and $B$ exhibits appropriate coincidences, the tiling is limit-periodic~\cite{Dekking1978,Queffelec1995}.  For the substitution matrix $(1,1,2,2)$,  the rule $[S\rightarrow SL$; $L\rightarrow SLSL]$ produces a periodic tiling, and $[S\rightarrow SL$; $L\rightarrow SLLS]$, for example, produces a limit-periodic tiling.   

For the limit-periodic cases, the analysis above would suggest $\alpha \rightarrow \infty$, or, more properly, $\alpha$ is not well defined.   We present here an analysis a particular case for which the convergence of $Z(k)$ to zero is indeed observed to be faster than any power law.  

\begin{figure}[h]
  \centering
  \includegraphics[width=\columnwidth]{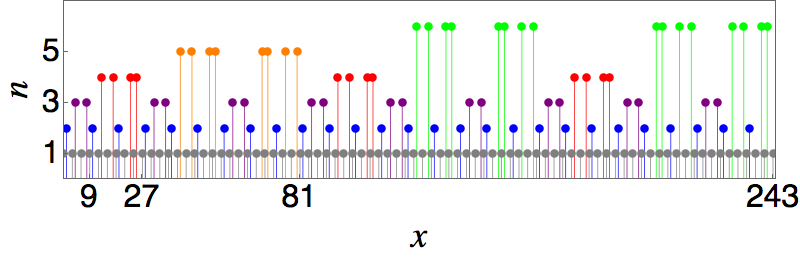}
  \includegraphics[width=\columnwidth]{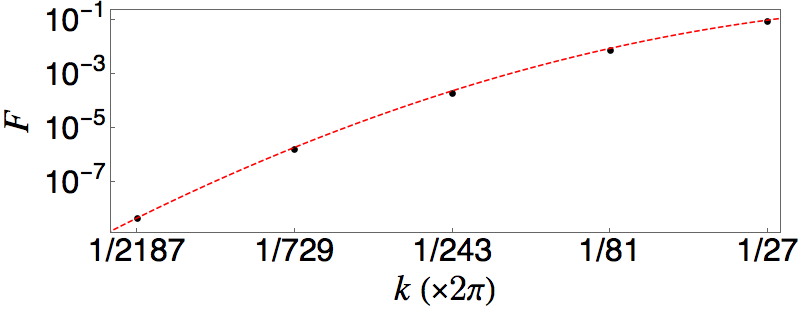}
  \includegraphics[width=\columnwidth]{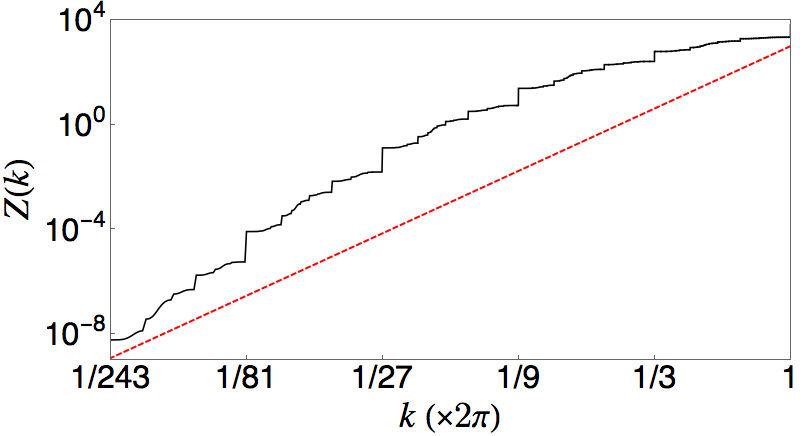}
  \includegraphics[width=\columnwidth]{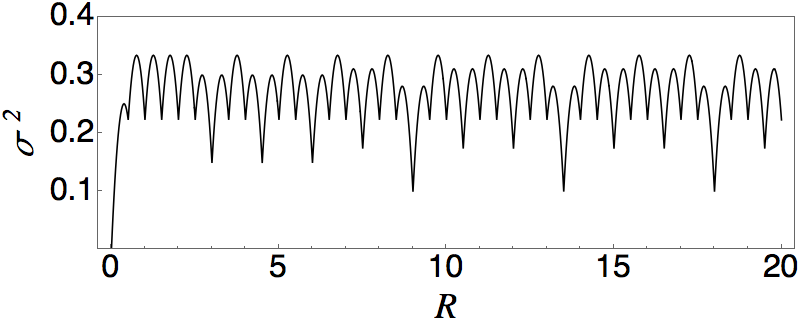}
  \caption{Top: Periodic sublattices of the limit-periodic point set generated by Eq.~(\ref{eqn:lprule}).  Each point is plotted at a height $n$ corresponding to the subset that contains it.  Points of the same color form a periodic pattern with period $3^n$.  Second: Deviation of $|A(k_n)|$ from $1/3^n$.  Third: The integrated structure factor for the limit-periodic tiling with $\lambda_2 = 0$, computed from subsets with $n\le 8$.  The straight red (dashed) line of slope 5 is a guide to the eye for observing the concavity of the curve.  Bottom: Plot of the number variance for the limit-periodic tiling with $\lambda_2 = 0$.}
\label{fig:lp0}
\end{figure}

The substitution rule 
\begin{equation}
S\rightarrow SL, \quad L\rightarrow SLLS
\label{eqn:lprule}
\end{equation}
with $S=1$ and $L=2$ produces a limit-periodic tiling with $a=1$ and $p = 3$.  Inspection of the point set (displayed in Fig.~\ref{fig:lp0}) reveals that the number of points in the basis of each periodic subset for $n\ge 2$ is $2^{n-2}$.  
The density of points in subset $n\ge 2$ is $(1/4)(2/3)^n$. 
The substitution matrix $\Mm = (1,1,2,2)$ has eigenvalues $\lambda_1 =3$ and $\lambda_2 = 0$.  

The unusual scaling in this case arises from interference effects associated with the form factors of the different periodic subsets.
Let $k_n = 2\pi/3^n$, the fundamental wavenumber for the $n$th subset, and let $X_n$ denote the set of points in a single unit cell of the $n$th subset.
$S(k_n)$ has contributions coming from all subsets of order $n$ and higher.  (Subsets of lower order do not contribute, as their fundamental wavenumber is larger than $k_n$.)  After some algebra, we find 
\begin{align}
A(k_n) =  & \frac{1}{3^n} \sum_{x\in X_n} e^{2\pi i x/3^n} \\
\ & + \frac{1}{3^{n+1}}\bigg[ \sum_{x\in X_{n+1}}\!\!\! e^{2\pi i x/3^n}  + \sum_{x\in X_{n+2}}\!\!\! e^{2\pi i x/3^n} \bigg].\nonumber
\end{align}
Numerical evaluation of the sums over the unit cell bases reveals that $A(k_n)$ is suppressed by the interference from subsets of higher order.  Figure~\ref{fig:lp0} shows the behavior of the quantity $F_n=3^n |A(k_n)|$, revealing a rapid decay for small $k_n$.  The red (dashed) line shows the curve $F_x = (1/3) (3x)^{-\ln(9x)/2}$, which appears to fit the points well.  

An analytic calculation of $Z(k_n)$ is beyond our present reach.  The middle panel of Fig.~\ref{fig:lp0} shows the results of a numerical computation that includes all peaks $k = 2\pi m/p^6$, with $p=3$.  It is clear that $Z(k)$ is concave downward on the log-log plot, consistent with the expectation that $Z(k)$ goes to zero faster 

\onecolumngrid

\begin{table*}[h]
\begin{tabular}{|c|c|c|c|c|c|c|}
\hline
\  & Anti-  & Weakly & Logarithmically & \multicolumn{3}{c|}{Strongly hyperuniform} \\ 
\  & hyperuniform  & hyperuniform  & hyperuniform & \multicolumn{3}{l|}{\ } \\ 
\  &  \  & (Class III) & (Class II) & (Class I) & Anomalous & Gapped \\  \hline
\  & $-1\leq\alpha \leq 0$ & $0 < \alpha < 1$ & $\alpha = 1$ & $1<\alpha \leq 3$ & $\alpha \rightarrow \infty$ & $\alpha$ irrelevant \\ \hline
Periodic & \No & \No & \No & \No & \No & \Yes \\
Quasiperiodic & ?  & ?  & \Yes & \Yes & ? & \No \\
Non-PV & \Yes & \Yes & \No & \No & \No & \No \\
Limit-periodic & \Yes & \Yes & \Yes & ? & \Yes & \No \\ \hline
\end{tabular}
\caption{Types of 1D tilings and their possible hyperuniformity classes.  A bullet indicates that tilings of the given type exist, a dash that there are no such tilings, and a question mark that we are not sure whether such tilings exist.  
}
\label{tab:types}
\end{table*}

\twocolumngrid

than any power of $k$.  Note that the curve is not reliable for the smallest values of $k$ due to the cutoff on the resolution of $k$ values that are included.  The deviation from power law scaling is most easily seen in the increasing with $n$ of the step sizes of the large jumps at $k = 2\pi/3^n$.  (Compare to the constant step sizes in Figs.~\ref{fig:lpdoubling}, \ref{fig:lppoisson}, and~\ref{fig:lpantihyper}.)

For completeness, the bottom panel of Fig.~\ref{fig:lp0} also shows a plot of the number variance for this tiling.  As expected, $\sigma^2(R)$ is bounded from above.  We note that the curve appears to be piecewise parabolic, which is also the case for the standard Fibonacci quasicrystal~\cite{Oguz2016}, though the technique for calculating $\sigma^2(R)$ based on the projecting the tiling vertices from a 2D lattice is not applicable here.

\section{Discussion}
\label{sec:discussion}
We have presented a heuristic method for calculating the hyperuniformity exponent $\alpha$ characterizing point sets generated by substitution rules that preserve the length ratios of the intervals between points.  The calculation relies only on the relevant substitution matrix and an assumption that the tile order under substitution does not lead to a periodic tiling.  The method performs well in that it yields a value of $\alpha$ consistent with direct measurements of the scaling of $\sigma^2(R)$ in several representative cases.  This allows for a straightforward construction of point sets with any value of $\alpha$ between $-1$ and $3$.

It is well known that substitution rules can be divided into distinct classes corresponding to substitution matrices having eigenvalues that are not PV numbers lead to structure factors $S(k)$ that are singular continuous~\cite{Bom86,Baake2017}, while substitution rules for which $|\lambda_2| < 1$ yield Bragg peaks.  Our analysis shows that this distinction corresponds to $\alpha$ greater than or less than unity, respectively.  From the perspective of hyperuniformity, on the other hand, the critical value of $\alpha$ is zero, which corresponds to $|\lambda_2|=\sqrt{\lambda_1}$.  To achieve $\alpha<0$, a naive comparison to scaling theories for systems with continuous spectra would suggest that $S(k)$ must diverge for small $k$.  We find, however, that anti-hyperuniformity, which does require sub-linear scaling of $Z(k)$, can occur without any divergence both in cases where the spectrum is singular continuous, as for non-PV substitutions, and in cases where the spectrum consists of a dense set of Bragg peaks, as in some limit-periodic systems.

Finally, our investigations led us to consider the results of applying substitution rules for which $\lambda_2 = 0$, which turned up a novel case of a limit-periodic tiling for which $S(k)$ approaches zero faster than any power law.  The physical implications of this type of scaling have yet to be explored.

The different tiling types and their hyperuniformity properties are summarized in Table~\ref{tab:types}.  Examples of quasiperiodic tilings in Classes~I and~II are presented in Ref.~\cite{Oguz2016}.  Note, however, that the Class~II case is not a substitution tiling.  We do not know whether some other construction methods might yield quasiperiodic tilings that are in Class~III, anti-hyperuniform, or even anomalous.   For non-PV tilings (which are substitution tilings by definition), at least two eigenvalues of the substitution matrix must be greater than unity, which rules out Class~II and Class~I.  We conjecture that there are no limit-periodic tilings in Class~I.   We can prove this for $D=2$ substitutions based on the fact that limit-periodicity requires the two eigenvalues to be rational and the fact that $\Mm$ has only integer elements requires their sum and product to be integers, but we do not have a proof for $D > 2$. 

\section{Acknowledgements}
J.E.S.S.~thanks Michael Baake, Franz G{\"a}hler, Uwe Grimm and Lorenzo Sadun for helpful conversations at a workshop sponsored by the International Centre for Mathematical Sciences in Edinburgh.  P.J.S. thanks the Simons Foundation for its support and New York University for their generous hospitality during his leave at the Center for Cosmology and Particle Physics where this work was completed.  S.T. was supported by the National Science Foundation under Award No.~DMR-1714722.
  
\appendix*
\section{Calculation of $\sigma^2(R)$ for the period doubling limit-periodic tiling}
The substitution rule $L\rightarrow LSS$, $S\rightarrow L$ (with $S=1$ and $L=2$) applied to an initial $L$ with its left boundary at $x=1$ produces a tiling with tile boundaries at all positions of the form $x_{m,j} = (2i + 1)4^m$, with $j$ and $m$ both running over all positive integers (including zero).  We are interested computing $\sigma^2(R)$ for the density 
\begin{equation} \label{eqn:rho1}
\rho(x) = \sum_{m,j = 0}^{\infty} \delta\big(x - (2j + 1)4^m\big)\,.
\end{equation}
Recall that $\sigma^2(R)$ is the variance in the number of points covered by a window of length $2R$ placed with it left edge at $x$ with uniform probability over all positive real values of $x$.

In Ref.~\cite{Torquato2018b}, an expression for $\sigma^2(R)$ is derived using Eq.~(\ref{eqn:local}) above.  Here we show how $\sigma^2(R)$ can be computed directly, thereby confirming the validity of Eq.~(\ref{eqn:local}) for this limit-periodic system and arriving at a particularly simple expression that can be analyzed in detail.

We first note that we can rewrite $\rho(x)$ as follows:
\begin{equation} \label{eqn:rho2}
\rho(x) = \sum_{n,i = 0}^{\infty} (-1)^n \delta\big(x - i 2^n\big)\,.
\end{equation}
To see this, first note that if $x$ is an odd integer, then the only term that contributes is $n=0$, $i=x$, which gives a $+1$.  This is the $m=0$ lattice of Eq.~(\ref{eqn:rho1}).  More generally, if $x$ is an odd multiple of $2^p$, there are contributions only from all $n \leq p$, and these have alternating signs.  If $p$ is odd, the number of such contributions is even, yielding a density of zero.  If $p$ is even, the sum of the contributions is $+1$.  The even values of $p$ correspond to the values of $m$ in Eq.~(\ref{eqn:rho1}).

Let $w = 2R$ be the length of the window, and let $N_n(x)$ be the number of points in the $n^{th}$ lattice covered by the window.  It is convenient to take the window to be open at its left edge and closed at its right edge.  Define $\{w\}_n$ as the fractional part of $w/2^n$.  It is convenient to think of $w$ as being expressed in base 2.  $\{w\}_n$ is then given by the first $n$ digits to the left of the decimal point, plus all of the digits to the right.  Note that $N_n(x)$ depends on $x$ only through $\{w\}_n$; the integer part of $w/2^n$ adds the same number of points independent of the value of $x$.  Furthermore, there are only two possible values of $N_n(x)$, which differ by unity.  For the purpose of computing the variance, we take these to be $0$ and $1$, and we work with the densities of these values rather than the full values of $N_n$.

If the window is placed with its left edge at $x = j 2^n$, the contribution to the density from the $n^{th}$ lattice is $0$.  In order for the window to cover an additional point, the left edge must be placed such that $\{x\}_n > 1 - \{w\}_n$.  The average density covered by the window of length $w$ is thus
\begin{equation}\label{eqn:rhoavg}
\langle\mu\rangle = \sum_{n=0}^{\infty} (-1)^n \{w\}_n\,.
\end{equation}

\begin{figure}
\includegraphics[width=\columnwidth]{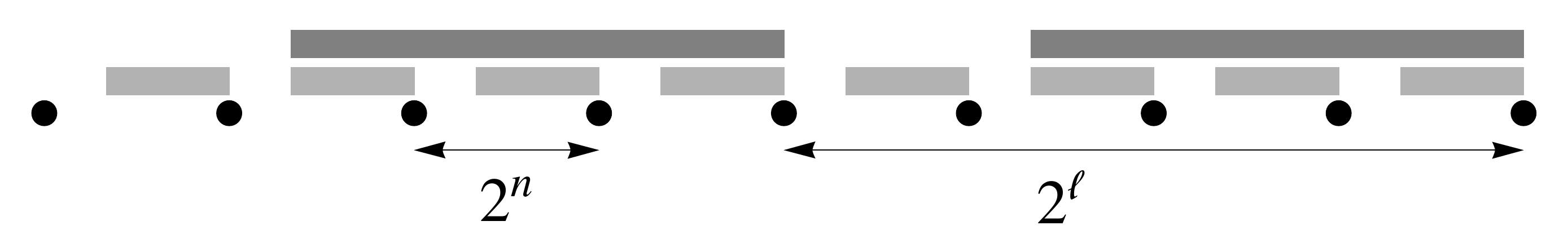}
\caption{Illustration for explaining the computation of a term in the double sum expression for $\langle\mu^2\rangle$.}
\label{fig:bars}
\end{figure}

The density squared is
\begin{equation} \label{eqn:rho2}
\rho^2(x) = \sum_{n,i = 0}^{\infty}\sum_{\ell,j = 0}^{\infty}  (-1)^{n+\ell} \delta\big(x - i 2^n\big) \delta\big(x - j 2^{\ell}\big)\,.
\end{equation}
A nonzero contribution to $\langle\mu^2\rangle$ arises from an individual term if and only if both the $n$ and $\ell$ lattices contribute.  For $\ell > n$, the fraction of $x$'s for which this is true is
\begin{equation}
\{w\}_n\left(\{w\}_{\ell} + 2^{n-\ell} (1-\{w\}_{n})\right)\,.
\end{equation}
The first term accounts for window placements that give a contribution from the $n^{th}$ lattice.  The light gray bars in Fig.~\ref{fig:bars} show the values of $x$ where the left edge of the window can be placed to produce a nonzero contribution.   The term in parentheses counts the number of such intervals that occur within a region that contribute from the $\ell^{th}$ lattice (indicated by dark grey bars in the figure) divided by the number of bars in one lattice spacing of the $\ell^{th}$ lattice.
We thus have
\begin{align}
\langle\mu^2\rangle = & \sum_{n = 0}^{\infty} \{w\}_n (-1)^{2n} \\ 
\ & + 2 \sum_{n = 0}^{\infty} \sum_{\ell > n}^{\infty} \{w\}_n\left(\{w\}_{\ell} + 2^{n-\ell} (1-\{w\}_{n})\right)\,. \nonumber
\end{align}
Using $(-1)^{2n} = 1$ and writing out $\langle\mu\rangle^2$ as a sum over $n$ plus a double sum with $\ell > n$, straightforward algebra with convenient cancellations of some of the double sums yields
\begin{eqnarray}
\sigma^2 & =  & \langle\mu^2\rangle - \langle\mu\rangle^2 \\
\ & = & \sum_{n=0}^{\infty} \big(\{w\}_n - \{w\}_n^2\big) \\ \nonumber 
\ & \ & + 2 \sum_{n=0}^{\infty}\sum_{\ell=n+1}^{\infty} \big(\{w\}_n - \{w\}_n^2\big)\left(\frac{-1}{2}\right)^{\ell - n} \\
\ & = & \frac{1}{3} \sum_{n=0}^{\infty}  \big(\{w\}_n - \{w\}_n^2\big)\,. \label{eqn:s2}
\end{eqnarray}
This result has been confirmed to be in perfect agreement with direct computations.

Equation~(\ref{eqn:s2}) describes a piecewise quadratic function of $w$.  (See Fig.~\ref{fig:lpdoubling}.) One immediately sees that all values of $w$ of the form $2^{\ell}$ give the same result; they give $\{w\}_n = 0$ for $n\leq \ell$ and the same infinite series for $n > \ell$.  Recalling that $R = w/2$, the shared value is
 \begin{equation}
\sigma^2(2^{n-1})  = \frac{1}{3}\sum_{m=1}^{\infty} \left(\frac{1}{2}\right)^m -  \left(\frac{1}{4}\right)^m = \frac{2}{9}\,.
 \end{equation}

To show that the upper envelope of $\sigma^2$ grows logarithmically, we first prove an upper bound that grows only logarithmically, then identify a special sequence of window length values for which the growth is logarithmic.  The upper bound is obtained by replacing all terms $\{w\}_n - \{w\}_n^2$ in the sum with the maximum value $1/4$ for all $n < 1+ \log_2 d$, then replacing the remaining infinite series with its maximal value, obtained by maximizing $\sum_n (x/2^n - x^2/4^n)$.  The result is 
\begin{equation}
\sigma^2(2^{n-2}<R<2^{n-1}) < \frac{1}{12}(3+n)\,.
\end{equation}

To show that there is a sequence of $R$ values for which $\sigma^2$ grows logarithmically, consider $w$ of the form $2^n/3$.  Note that the binary representation of $w$ is $101\ldots01.0101 \ldots$ for $n$ odd and $\ldots 101\ldots0.1010 \ldots$ for $n$ even.  Again we consider the contributions from $\ell \leq n$, then sum the remaining series.  The value of $\{w\}_{\ell}$ oscillates between $1/3$ and $2/3$ for $\ell<n$.  Straightforward algebra yields
\begin{equation}
\sigma^2(R = 2^{n}/3) = \frac{2n}{27} + \frac{26}{81}\,,
\end{equation}
which clearly grows logarithmically with $R$.  Note that the coefficient of $n$ here is $2/27$, reasonably close to the coefficient of $1/12$ derived for the upper bound, and that, as shown in Fig.~\ref{fig:lpdoubling}, these points are quite close to the true maxima for $w<2^n$.

\bibliographystyle{unsrt}
\normalbaselines
\bibliography{../hu}

\end{document}